# Synaptic modulation of conductivity and magnetism in a CoPt-based electrochemical transistor

*Shengyao Li,[1][†] Bojun Miao,[1][†] Xueyan Wang,[1] Siew Lang Teo,[2] Ming Lin,[2] Qiang Zhu,[2] S. N. Piramanayagam,[1][*] X. Renshaw Wang[1,3][*]*

[†]These authors contribute equally: Shengyao Li, Bojun Miao
[*]These authors are co-corresponding authors: X. Renshaw Wang, S. N. Piramanayagam

Shengyao Li, Bojun Miao, Xueyan Wang, S. N. Piramanayagam, X. Renshaw Wang
School of Physical and Mathematical Sciences, Nanyang Technological University, 637371, Singapore
Email: renshaw@ntu.edu.sg
Email: prem@ntu.edu.sg

X. Renshaw Wang
School of Electrical and Electronic Engineering, Nanyang Technological University, 639798, Singapore

Siew Lang Teo, Ming Lin, Qiang Zhu
Institute of Materials Research and Engineering (IMRE), A*STAR, 138634, Singapore

Abstract: Among various types of neuromorphic devices towards artificial intelligence, the electrochemical synaptic transistor emerges, in which the channel conductance is modulated by the insertion of ions according to the history of gate voltage across the electrolyte. Despite the striking progress in exploring novel channel materials, few studies report on the ferromagnetic metal-based synaptic transistors, limiting the development of spin-based neuromorphic devices. Here, we present synaptic modulation of both conductivity as well as magnetism based on an electrochemical transistor with a metallic channel of ferromagnetic CoPt alloy. We first demonstrate its essential synaptic functionalities in the transistor, including depression and potentiation of synaptic weight, and paired-pulse facilitation. Then, we show a short- to long-term plasticity transition induced by different gate parameters, such as amplitude, duration, and frequency. Furthermore, the device presents multilevel and reversible nonvolatile states in both conductivity and coercivity. The results demonstrate simultaneous modulation of conductivity and magnetism, paving the way for building future spin-based multifunctional synaptic devices.

**1. Introduction**

Neuromorphic computing, which emulates the working principle of biological neurons and synapses, provides unique computing and recording capabilities that could break the limitation of conventional von Neumann computing.[1] In a biological system, a synapse is a critical joint connection between the pre- and post-synaptic cells, and its connection can be strengthened by the guided migration of neurotransmitters.[2],[3] Inspired by this biological structure, electronic synaptic devices with similar tunable connections were developed towards neuromorphic computing. Out of many types of synaptic devices, the electrochemical synaptic transistors based on electrolytes have attracted much attention in recent years.[4] In these transistors, ions are driven in and out of the channel with the application of gate voltage ($V_G$) across the electrolyte, inducing multilevel and nonvolatile conductance changes in the channel. Over the past years, a variety of channel materials have been investigated, including van der Waals materials,[5]–[9] metal oxides,[10]–[16] and organic materials.[17] Despite the progress in electrochemical synaptic transistors, few studies report the ferromagnetic metal-based synaptic transistors, especially the synaptic modulation of magnetism upon ion intercalation.

The CoPt alloy is of particular interest to be used as a channel material of a synaptic transistor owing to its high conductivity and spin degree of freedom.[18] The perpendicular magnetization of the CoPt film gives rise to the anomalous Hall effect (AHE) and possibly spin-orbit torque (SOT) switching.[19],[20] Moreover, the CoPt alloy preserves the perpendicular magnetization even in thick films, providing thermal stability and eventually device reliability.[21],[22] Theoretically, the perpendicular magnetization can be tuned by the ion insertion in an electrochemical transistor, resulting in variations in coercivity ($H_C$), saturation magnetization and Curie temperature.[23]–[25] The toggle of perpendicular magnetization is significantly useful to spintronic devices. Therefore, the technical advantages of perpendicular magnetization, high conductivity, and stability position the CoPt alloy-based electrochemical transistor as one of the ideal candidates for achieving spin-based synaptic devices.

In this study, we have fabricated a CoPt-based electrochemical transistor and emulated the biological synapse. The sputter-grown Ta (2 nm)/ $Co_{80}Pt_{20}$ (9 nm) channel on the $SiO_2$/Si substrate with a perpendicular magnetization presents an AHE loop under an out-of-plane magnetic field. The device successfully emulates the functionality of a biological synapse, including depression and potentiation of synaptic weight, paired-pulse facilitation (PPF), and the short-term plasticity (STP) to long-term plasticity (LTP) transition, which can be tuned by gate parameters such as the pulse number, duration, frequency and amplitude. Moreover, both the conductivity and the magnetic $H_C$ present multilevel, nonvolatile, and reversible transitions according to the history of $V_G$, which also has an excellent retention characteristic.

## 2. Results and discussion

**Figure 1**a schematically illustrates how a synaptic transistor can emulate a biological synapse. The gate electrode, ionic liquid and channel can work as a presynaptic neuron, neurotransmitter, and postsynaptic neuron, respectively. The analogue signal transmission in the biological synapse is realized by releasing and receiving neurotransmitters between synaptic structures. While in electrochemical transistors, it is realized by the ion migration under the voltage across the ionic liquid. With the application of $V_G$, the cations and anions accumulate at the channel/IL and gate/IL interfaces, respectively, forming an electric double layer (EDL) and contributing to the volatile modulation of the conductance. On the other hand, once the $V_G$ exceeds the threshold voltage of the ion insertion, partial ions are trapped in the channel, leading to the nonvolatile modulation of conductance.[11] After the removal of $V_G$, ions gradually retreat to the ionic liquid due to the concentration gradient, and the channel's ability to preserve the ions determines the retention property.[26]

Figure 1b shows the optical image of our device. The channel consists of Ta (2 nm)/ CoPt (9 nm) stack on (285 nm) $SiO_2$/Si substrate, which exhibits a strong perpendicular magnetization (Figure S1). A small drop (10 μL) of ionic liquid (IL) *N*,*N*-diethyl-*N*-methyl-*N*-(2-methoxyethyl)ammonium bis(trifluoromethanesulfonyl)imide (DEME-TFSI) was applied and covers both the gate electrode and the channel. In this structure, the conductance and magnetism of the CoPt channel could be modulated by the intercalation of ions upon the application of $V_G$. More details of the device fabrication can be found in the Experimental Section.

A nonvolatile post-synaptic conductance (PSC) after a voltage stimulation is crucial for emulating a biological synapse. The excited PSC is termed the excitatory PSC (EPSC), while inhibited PSC is termed the inhibitory PSC (IPSC).[27] Synaptic plasticity can be further divided into long-term plasticity (LTP) and short-term plasticity (STP) depending on the retention capability of the PSC, which plays an essential role in mimicking the learning, memory, and behavioural adaptation in the biological synapse.[28] To achieve the PSC, the $V_G$, which is an external stimulus, was applied between the gate electrode and drain electrode (see Figure 1b).

Figure 1c shows the EPSC (red line) and IPSC (blue line) by applying a $V_G$ of 2 and -2 V for 30 s. At the rising edge of each $V_G$, the conductance begins to increase (decrease) for the positive (negative) $V_G$, gradually reaches the highest (lowest) value, and eventually decays to a certain level, resulting in a nonvolatile PSC.

The channel conductance as a function of $V_G$ (Figure 1d) was characterized by anticlockwise sweeping the $V_G$ at various speeds ranging from 0.005 to 0.15 V/s. The transfer curve loops of the electrochemical transistor are enhanced under a slower sweeping rate, indicating a larger modulation rate due to enhanced ion intercalated at a slower sweeping rate. Figure 1e shows the leakage current ($I_G$) as a function of $V_G$. The $I_G$ is negligible (~0.1 nA) compared with the input current in the channel (10 µA). Therefore, the ionic liquid is ionically conducting but electrically insulating, and it does not affect the result of conductance modulation.

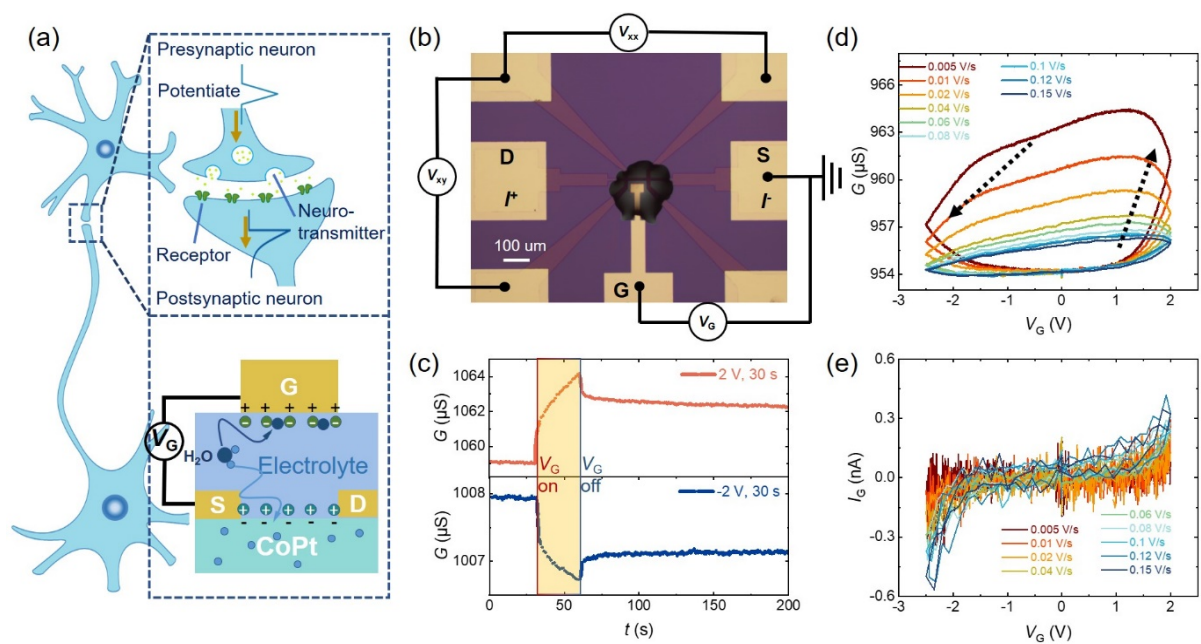

**Figure 1.** Schematic diagram and nonvolatile property of the synaptic device based on CoPt. a) Schematic illustration of a biological axodendritic synapse and the analogue of a synaptic transistor. The plasticity of a biological synapse is modulated by the biological voltage between pre-and post-neurons, while the conductance of the CoPt channel is modulated by the ion insertion driven by the gate voltage ($V_G$). b) Optical image of the CoPt alloy-based synaptic transistor, the current is applied between the source (S) and drain (D) electrodes, and the gate voltage is applied between D and gate (G) electrodes. c) Nonvolatile responses of channel conductance after single pulse stimulation of positive and negative voltages. Conductance potentiation (depression) occurs when the rising (falling) edge is triggered. d) Transfer curves

under anticlockwise sweeping gate voltages at different rates. e) Leak current ($I_G$) as a function of $V_G$ under different sweeping rates.

The ion insertion is activated when $V_G$ exceeds the threshold voltage, while the insertion rate can be tuned by varying the gate parameters, including the gate pulse amplitude, duration, and number. **Figure 2**a shows the EPSC realized by applying a series of $V_G$ ($t$ = 30 s) with amplitudes ranging from 1.2 to 2.6 V. More EPSC after the application of different pulse duration and numbers, as well as the IPSC results, are shown in Figure S4. The EPSC peak value and the LTP rises with the increase of pulse amplitude, duration, and number. The conductance at the peak and 300 s after the removal of $V_G$ are defined as the volatile and nonvolatile modulation of the conductance, respectively. Figure 2b shows the net conductance modulation ($\Delta G$), which is defined as the conductance variation against that at the beginning of the measurement. $\Delta G$ increases with the increase of pulse amplitude, especially for the nonvolatile region of the conductance modulation. The results indicate that higher $V_G$ promotes the ion insertion into the CoPt channel and contributes to the nonvolatile change of the conductance of the CoPt channel.

The perpendicular magnetic anisotropy and the resultant modulation of our sputter-deposited CoPt alloy were electrically characterised by measuring its AHE loops after the application of various $V_G$.[18],[29],[30] Figure 2c shows the widening of the AHE loops of the Hall resistance ($R_{xy}$) after the application of a series of negative $V_G$ for 60 s. The $H_C$ of the AHE loops is remarkably widened under a larger $V_G$ (see Figure 2e). The results indicate that the ion insertion produces a similar effect on the magnetic $H_C$ as that on the conductance. Therefore, the CoPt-based synaptic transistor demonstrates multiple magnetic states controlled by the ion insertion, which set it apart from conventional synaptic transistors.

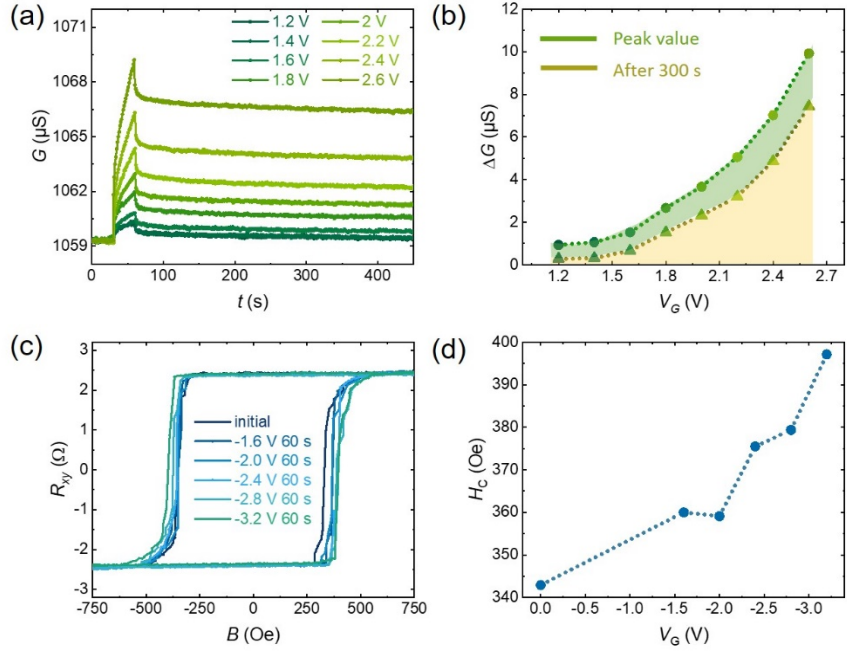

**Figure 2.** Synaptic potentiation on both conductivity and magnetic coercivity ($H_C$). a,b) Single pulse modulation of conductance under different gate amplitudes ($V_G$ = 1.2 V to 2.6 V, duration is 60 s). Even though the instantaneous depletion phenomenon at the falling edge leads to a certain conductance loss of modulation, its nonvolatile conductance still positively correlates with the amplitude of $V_G$ and reaches a stable state of more than 400s after the modulation. c,d) Single pulse modulation of AHE loops under different amplitudes ($V_G$ = -3.2 to -1.6 V, duration is 60 s), and the magnetic $H_C$ is increased after a larger $V_G$.

In biological synapses, paired-pulse facilitation (PPF) is an important form of STP, which is essential in processing auditory or visual signals based on short-term synaptic plasticity.[31],[32] PPF behaviour was investigated in our CoPt-based synaptic transistor by applying a pair of pulses with different intervals. **Figure 3**a inset shows the response of channel conductance to a pair of gate pulses with a magnitude of 2 V, duration of 1 s, and interval of 7 s. Two peaks of conductance are evoked by the voltage stimulation, the second peak is 1.24 times higher than the first peak due to the residue ions trapped in the channel. The PPF index is defined as the equation

$$PPF = 100\% \cdot A_2/A_1 , \qquad (1)$$

where $A_1$ and $A_2$ are the height of the first and second peaks, respectively. The green dots in Figure 3a displays the PPF index as a function of the pulse interval with amplitude and duration fixed at 2 V and 1 s, respectively. We obtain the highest PPF index of 135% at the smallest interval (1 s) because of a larger amount of residue ions in the channel. Then, the PPF variation is fitted with a standard PPF equation

$$PPF = 1 + C_1 \cdot \exp(-\Delta t/\tau_1) + C_2 \cdot \exp(-\Delta t/\tau_2), \qquad (2)$$

where $C_1$ and $C_2$ are the magnitudes of the initial facilitation, $\tau_1$ and $\tau_2$ are the characteristic relaxation times of the slow and rapid decay phases, respectively. The PPF index can be well fitted by the orange dash line, yielding a rapid ($\tau_1$ = 3.1 s) and slow ($\tau_2$ = 700 s) phases. In contrast, the rapid ($\tau_1$ = 2.8 s) and slow ($\tau_2$ = 700 s) phases are obtained by fitting the paired-pulse depression (PPD) index as a function of pulse interval (see Figure S6). Therefore, both the PPD and PPF demonstrate a double-exponential decay, which is similar to those biological synapse behaviours.[33] To emulate a repetitive learning process, which is equivalent to the application of a series of pulse stimuli, we applied 30 consecutive positive gate pulses (2 V, 2 s) spaced apart by 2 s. The conductance peak values of each stimulation ($A_1$ to $A_{30}$) are extracted in Figure 3b, showing an exponential growth tendency with the increase in pulse number.

Furthermore, thanks to the slow diffusion rate of the ions in the electrolyte, the electrochemical transistor can also serve as a high pass filter for sophisticated information processing. To characterize the frequency-dependent conductance change, a series of 5 pulse stimuli (2 V, 1 s) with varying frequencies from 0.09 to 0.5 Hz were applied to emulate the signal filtering characteristics of the device (Figure 3c). The amplitude gain ($A_5/A_1$) increases as a function of frequency, proving the high pass filter characteristic of the device (Figure 3d).

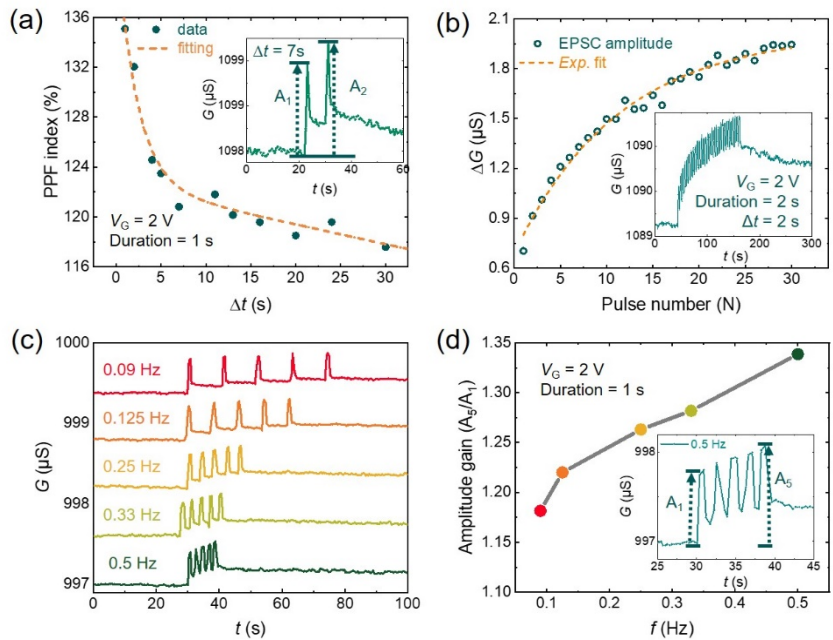

**Figure 3.** Frequency-dependent potentiation of CoPt-based electrochemical transistor. a) PPF index as a function of the pulse interval (green dots). The PPF index obeys the double exponential decay with time intervals. b) The peak value of conductance, which was evoked by each pulse, rises exponentially as a function of pulse number. The inset is the EPSC with the application of 30 consecutive pulses. c) EPSC stimulated by 5 consecutive pulses (2 V, 1 s) at different frequencies from 0.09 to 0.5 Hz. d) Amplitude gain ($A_5/A_1$) as a function of frequency. The inset shows that the $A_1$ and $A_5$ are the peak values of conductance modulation evoked by the first and fifth pulse, respectively.

We now characterise the multilevel, nonvolatile, and reversible states of the CoPt-based electrochemical transistor. To demonstrate the multilevel states, 7 continuously increasing gate pulses (from 1.4 to 2.6 V, duration of 30 s) are applied (**Figure 4**a). The EPSC amplitude gradually increases and the channel conductance shows a multilevel nonvolatile behaviour, with the conductance quickly reaches to a stable value shortly after the gate pulse. 10 negative pulses (-2.4 V, duration of 30 s) are then applied. The IPSC gradually decreases, and also shows a multilevel nonvolatile behaviour. The nonvolatile conductance after each gate voltage potentiation (depression) is extracted in Figure 4b, demonstrating a reversible multilevel conductance modulation. The conductance versus $V_G$ amplitude shows an exponential rising tendency, while the conductance versus gate pulse number is linearly related.

Moreover, the multilevel modulation of magnetic $H_C$ is also reversible (see Figure 4c). After the application of 4 continuously decreasing negative gate pulses (from -2 to -3.2 V, duration of 60 s), the AHE loops are widened. In contrast, the AHE loops are subsequently narrowed by applying 4 continuously increasing negative gate pulses (from 1.4 to 2.6 V, duration of 60 s). The magnetic $H_C$ of each AHE loop is extracted in Figure 4d. Five levels can be distinguished, with the magnetic $H_C$ being enlarged under negative $V_G$ and decreased under positive $V_G$. The multilevel reversible magnetic $H_C$ modulation suggests that the CoPt-based electrochemical synaptic transistor provides a platform for future spin-based neuromorphic devices. Combined with the domain wall (DW) devices or the spin-orbit torque (SOT) devices,[34]–[37] the switching current can be effectively controlled after sufficient training. This feature is applicable to in-memory computing, which is of high precision in deep learning.[38]

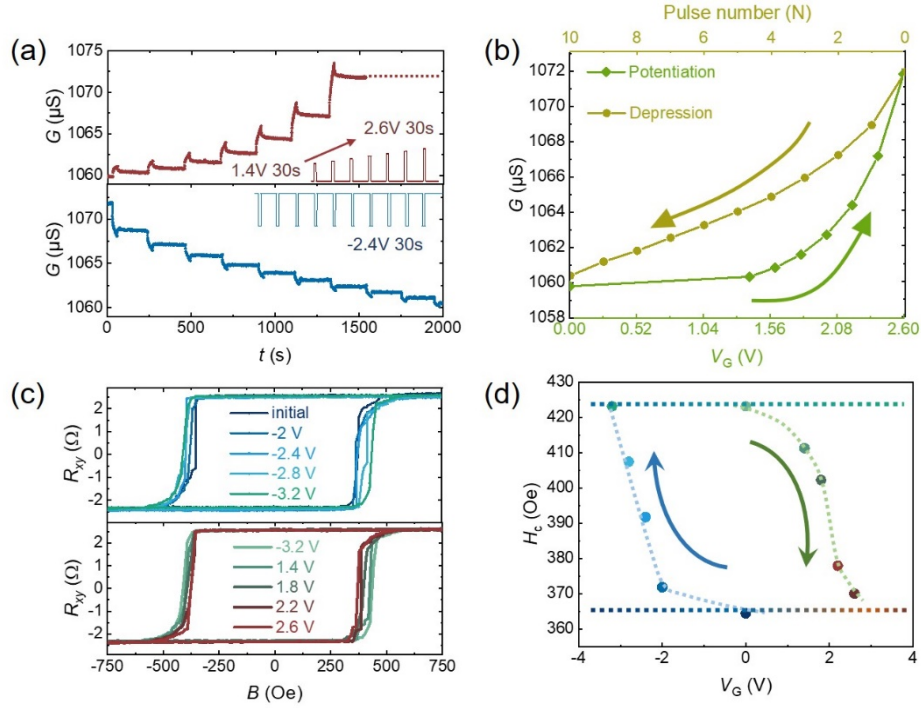

**Figure 4.** Multilevel and reversible modulation on conductance and magnetic $H_C$. a) Channel conductance modulation under 7 continuously increasing positive pulses (from 1.4 to 2.6 V, duration of 30 s), followed by 10 negative pulses (-2.4 V for 30 s). b) The nonvolatile, multilevel and reversible changes of conductance after each gate pulse stimuli. c) AHE loop modulation under 4 continuously decreasing negative gate voltages (from -2 to -3.2 V, duration of 60 s), followed by 4 continuously increasing positive voltages (from 1.4 to 2.6 V, duration of 60 s). d) Cyclical change of magnetic $H_C$.

To emulate the long-term repeatability of the conductance and magnetic $H_C$ potentiation and depression, we measured the cycle-to-cycle channel conductance variation (**Figure 5**a) by applying 20 consecutive positive gate pulses (1.8 V for 2 s, spaced 2 s apart) followed by 20 consecutive negative gate pulses (-2.4 V for 2 s, spaced 2 s apart). 5 cycles of the positive and negative pulses were applied, with a continuous conductance switch between the high conductance state (999.5 μS) and the low conductance state (997.8 μS), indicating good repeatability of synaptic long-term conductivity modulation (Figure 5b). Figure 5c displays the conductance variation as a function of time. After applying 50 consecutive pulses (2 V for 4 s spaced by 2 s), the high conductance state lasts for at least 2500 s, presenting a good retention characteristic after sufficient training.

Meanwhile, the electrochemical modulation of magnetic $H_C$ also demonstrates good repeatability. 5 cycles of $V_G$ (2.6 V for 60 s) and (-3.2 V for 60 s) are applied alternately, and the AHE loops were measured after the removal of $V_G$ (see Figure 5d). The AHE loops are sequentially widened (narrowed) after the application of negative (positive) $V_G$. Figure 5e shows that the magnetic $H_C$ alternates between the high state (400 Oe) and the low state (342 Oe). After the application of 50 consecutive pulses (-3.2 V for 4 s spaced by 2 s) to strengthen the transition from STP to LTP. Then, after the removal of $V_G$, the AHE loops do not show a variation for at least $10^4$ s, indicating good retention behaviour.

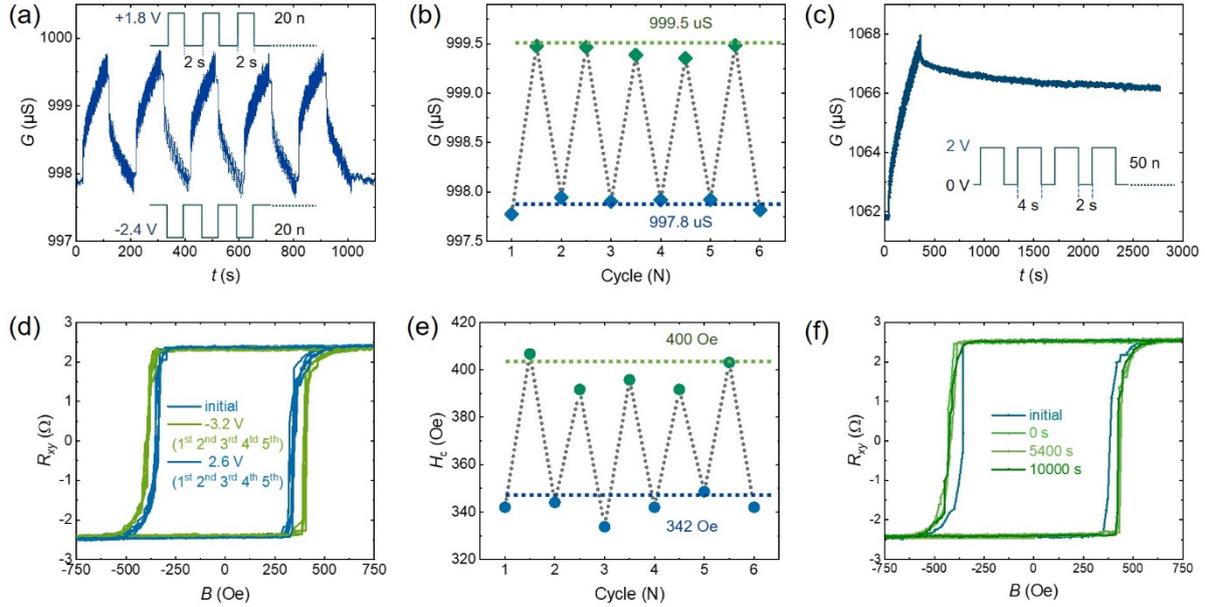

**Figure 5.** Long-term repeatability of the conductance and magnetic $H_C$ potentiation and depression. a) Changes in the channel conductance under 5 cycles of pulses, 20 consecutive positive pulses (1.8 V for 2 s spaced by 2 s) and 20 consecutive negative pulses (-2.4 V for 2 s spaced by 2 s) for one cycle. b) The modulation of conductance after 20 consecutive positive (negative) pulses. c) Retention behaviour of the EPSC after applying 50 consecutive positive gate pulses (2 V for 4 s spaced by 2 s). d) Changes in AHE loops under 5 cycles of gate modulation with a negative $V_G$ (-3.2 V, 60 s) and followed by a positive $V_G$ (2.6 V, 60 s) for one cycle. e) Evolution of $H_C$ after each gate stimulus. f) Retention behaviour of the AHE loop after applying 50 consecutive gate pulses (-3.2 V for 4 s spaced by 2 s).

## 3. Conclusion

In summary, we demonstrate an electrochemical synaptic transistor based on CoPt alloy. Multilevel, reversible and nonvolatile conductance and magnetic $H_C$ can be produced by controlling the intercalation and extraction of the ions according to the history of gate voltage. Based on the ion migration and insertion, the device successfully emulates the biological synaptic functionalities, including the STP to LTP transition, paired-pulse facilitation, and potentiation (depression) with different synaptic weights. Moreover, these properties are controllable by gate pulse amplitude, duration, and frequency. Both the conductance and magnetic $H_C$ present satisfying repeatability and retention properties. The results suggest the ferromagnetic metal-based electrochemical synaptic transistor is a promising candidate for spin-based synaptic transistors in future neuromorphic computing applications.

## 4. Experimental Section

### 4.1. Device Fabrication

Before the patterning, SiO$_2$/Si substrate was thoroughly cleaned in an ultrasonic bath with acetone and deionized water and dried by nitrogen stream. AZ 5214E photoresist was spin-coated at a speed of 4000 rpm for 45 s and prebaked on a hotplate at 110 °C for 90 s. Then, the channel was patterned into a Hall bar structure with a dimension of 80 μm ($L$) × 10 μm ($W$) using maskless UV lithography (TTT-07-UVlitho, TuoTuo Technology). Back sputtering was carried out at an Ar pressure of 10 mTorr and DC power of 40 W before the deposition to remove the photoresist residue. 2 nm Ta buffer layer was first deposited at an Ar pressure of 3 mTorr and using a DC power of 40 W. 9 nm Co$_{80}$Pt$_{20}$ was subsequently deposited at an Ar pressure of 20 mTorr and using a DC power of 40 W. The electrodes consist of Cr (2 nm)/Au (50 nm) were deposited using thermal evaporation. After liftoff with acetone and dried by nitrogen stream, a small drop of DEME-TFSI (Kanto Chemical) ionic liquid was applied on top of the device, covering the gate electrode and the channel. The device was then loaded into a chamber, and its vacuum of 10$^{-3}$ mbar was maintained using a scroll pump to protect the ionic liquid from the ambient moisture.

### 4.2. Electrical measurement

All the electrical measurements were performed at room temperature. A Yokogawa GS200 DC sourcemeter was used to supply $I_{SD}$ = 10 μA across the source (S) and drain (D) electrodes to avoid the heating effect of the channel. Four probe measurement was applied to eliminate the contacting resistance using a Keithley 2000 multimeter. A Keithley 6517B was used to apply the gate voltage and simultaneously measure the leakage current. All the AHE loops were measured immediately after the removal of $V_G$, with an out-of-plane magnetic field sweeping at 15 Oe/s.

**Acknowledgements**

The authors acknowledge the funding from the National Research Foundation (NRF), Singapore under its 21st Competitive Research Programs (CRP Grant No. NRF-CRP21-2018-0003). X.R.W. acknowledges support from Academic Research Fund Tier 2 (Grant No. MOE-T2EP50120-0006) from Singapore Ministry of Education, and the Agency for Science, Technology and Research (A*STAR) under its AME IRG grant (Project No. A20E5c0094). S. L acknowledges the research scholarship from CRP grant. X.R.W. and S.N.P. designed and directed this study.